\documentclass[twocolumn,showpacs,preprintnumbers,amsmath,amssymb,superscriptaddress,aps,prl]{revtex4}

\usepackage{graphicx}

\begin{document}

\title{Combination of Bloch oscillations with a Ramsey-Bord\'e interferometer : new determination of the fine
structure constant } 
\author{Malo~Cadoret}
\affiliation{Laboratoire Kastler Brossel, Ecole Normale Sup\'erieure, Universit\'e Pierre et Marie Curie, CNRS, 4 place Jussieu, 75252 Paris Cedex 05, France}
\author{Estefania~de~Mirandes}
\affiliation{Laboratoire Kastler Brossel, Ecole Normale Sup\'erieure, Universit\'e Pierre et Marie Curie, CNRS, 4 place Jussieu, 75252 Paris Cedex 05, France}
\author{Pierre~Clad\'e}
\affiliation{Laboratoire Kastler Brossel, Ecole Normale Sup\'erieure, Universit\'e Pierre et Marie Curie, CNRS, 4 place Jussieu, 75252 Paris Cedex 05, France}
\author{Sa\"\i da~Guellati-Kh\'elifa}
\affiliation{Conservatoire National des Arts et M\'etiers,
292 rue Saint Martin, 75141 Paris Cedex 03, France}
\author{Catherine~Schwob}
\affiliation{Laboratoire Kastler Brossel, Ecole Normale Sup\'erieure, Universit\'e Pierre et Marie Curie, CNRS, 4 place Jussieu, 75252 Paris Cedex 05, France}
\author{Fran\c cois~Nez}
\affiliation{Laboratoire Kastler Brossel, Ecole Normale Sup\'erieure, Universit\'e Pierre et Marie Curie, CNRS, 4 place Jussieu, 75252 Paris Cedex 05, France}
\author{Lucile~Julien}
\affiliation{Laboratoire Kastler Brossel, Ecole Normale Sup\'erieure, Universit\'e Pierre et Marie Curie, CNRS, 4 place Jussieu, 75252 Paris Cedex 05, France}
\author{Fran\c cois~Biraben}
\affiliation{Laboratoire Kastler Brossel, Ecole Normale Sup\'erieure, Universit\'e Pierre et Marie Curie, CNRS, 4 place Jussieu, 75252 Paris Cedex 05, France}

\begin{abstract}
We report a new experimental scheme which combines atom interferometry with Bloch oscillations to provide a new measurement
of the ratio $h/m_{\mathrm{Rb}}$. By using Bloch oscillations, we impart to the atoms up to $1600$ recoil momenta and  thus we improve the accuracy on the recoil velocity measurement. The deduced value of $h/m_{\mathrm{Rb}}$ leads to a new determination of the fine structure constant $\alpha^{-1}=137.035\,999\,45\,(62)$ with a relative uncertainty of $4.6\times 10^{-9}$. The comparison of this result with the value deduced from the measurement of the electron anomaly provides the most stringent test of QED.
\end{abstract}

\maketitle

The fine structure constant $\alpha$ characterizes the strength of the electromagnetic interaction. It is a corner stone of the adjustment of the fundamental constants \cite{codata06} and its determination is made in different domains of physics. The most precise values of $\alpha$ are deduced from the measurements of the electron anomaly $a_e$ made in the eighties by H.G.~Dehmelt at the University of Washington \cite{VanDick}, and, recently, by G.~Gabrielse at Harvard University \cite{Gabrielse2008}. This last measurement and an impressive improvement of quantum electrodynamics (QED) calculations \cite{Kinoshita} gives a value of $\alpha$ with a relative uncertainty of $3.7 \times 10^{-10}$ which surpasses the Dehmelt's result by one order of magnitude. Nevertheless this recent determination of $\alpha$ relies on very difficult QED calculations. To test them, other determinations of $\alpha$, independent of QED, are required. The most precise are deduced from the measurement of the ratio $h/m$ between the Planck constant and the mass of an atom thanks to the relation deduced from the ionization energy of hydrogen:
\begin{equation}
       \alpha^2=\frac{2R_\infty}{c}\frac{m}{m_e}\frac{h}{m},
\label{eqn1}
\end{equation}
where $m_e$ is the electron mass. The limiting factor is the ratio $h/m$: the uncertainty of the Rydberg constant $R_\infty$ is $7 \times 10^{-12}$ \cite{Schwob,Udem} and that of the mass ratio $m/m_e$ $4.8\times10^{-10}$ \cite{{codata06},{Bradley}}.
This principle has been used on Cs atom by S. Chu and colleagues with an atom interferometer to measure the recoil velocity $v_r=\hbar k /m$ of the atom when it absorbs a photon of momentum $\hbar k$ and to deduce $\alpha$ with a relative uncertainty of $7.4 \times 10^{-9}$ \cite{Wicht}. In 2005, our group used the Bloch oscillations (BO) of Rb atoms in an optical lattice to transfer a large number of photon momenta to the atoms: we achieved a precise measurement of atomic recoil velocity and a determination of $\alpha$ with a relative uncertainty of $6.7\times 10^{-9}$ \cite{{CladePRL},{CladePRA06}}.

In this letter, we present a combination of the two methods: we use BO to transfer a large number of photon momenta and a Ramsey-Bord\'e interferometer to precisely measure the induced atomic velocity variation. This leads to a value of $\alpha$ with a relative accuracy of $4.6\times 10^{-9}$. The comparison of this result, which is very slightly dependant on QED, with the value of $\alpha$ deduced from the electron anomaly is the most stringent test of the QED calculations.

Bloch oscillations have been first observed in atomic physics by the groups of C. Salomon and M.G. Raizen \cite{{Dahan},{Raizen}}. The atoms are placed in an optical lattice by shining on them two counter-propagating laser beams whose frequency difference is swept linearly. The atoms undergo a succession of Raman transitions which correspond to the absorption of one photon from a beam and a stimulated emission of a photon to the other beam. The internal state is unchanged while the atomic velocity increases by $2v_r$ per oscillation. The Doppler shift due to this velocity variation is periodically compensated by the frequency sweep and the atoms are accelerated. This method is very efficient to transfer to the atoms a very large number of photon momenta \cite{Battesti}. Another point of view is to consider that the atoms are placed in a standing wave which is accelerated when the frequency difference between the two laser beams is swept. In the atomic frame, the atoms undergo an inertial force. This system is analog to the BO of an electron in a solid submitted to an electric field \cite{Dahan}. There is a similar context when an atom is placed in a vertical standing wave in the gravitational field. This situation has been already investigated for measuring the local acceleration of gravity \cite{Ander,Roati, EPL06}. It has also been suggested to use BO to probe forces near surfaces at a microscopic scale \cite{Carusotto,Wolf}.

\begin{figure}
\centering
\includegraphics[width=\linewidth]{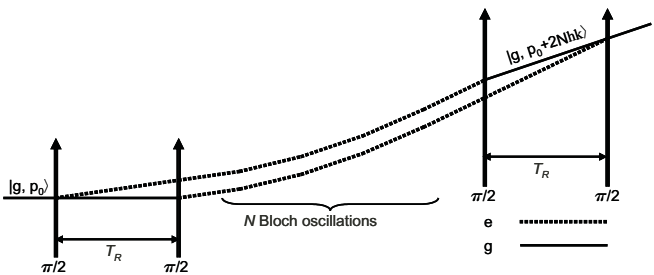}
\caption{\label{fig:Interferometer} Interferometric scheme combining
Ramsey-Bord\'e interferometer and $N$ Bloch oscillations.}
\end{figure}

In our experiment BO are combined with a Ramsey-Bord\'e interferometer \cite{Borde} following the scheme of Fig. \ref{fig:Interferometer}. BO are inserted between two pairs of $\pi/2$ laser pulses. Each pulse drives a Doppler-sensitive Raman transition between the two hyperfine ground states labeled $|g\rangle$ and $|e\rangle$ ($F=2$ and $F=1$ in the case of $^{87}$Rb). The first pair
of $\pi/2$ pulses creates the two coherent atomic wavepackets in internal state $|e\rangle$. We then blow away the atoms remaining in the initial state $|g\rangle$ by using a laser beam tuned to a single photon transition. Both wavepackets are accelerated by $N$ BO. The second $\pi/2$ pair recombines the two atomic wavepackets and readouts the phase difference between the two paths. The interferometer can also be understood in momentum space. After the first pair of $\pi/2$ pulses, the shape of the velocity distribution of the atoms in the level $|e\rangle$ is a Ramsey fringe pattern. The fringe spacing varies as  $1/T_R$ ($T_R$ is the delay between the two $\pi/2$ pulses). These atoms are adiabatically loaded into the first Brillouin zone of the optical lattice. Following the Bloch formalism, if the atom has a well defined momentum $\hbar q_0$ with $|q_0|<k$, the atomic wave function is modified when the optical potential is increased adiabatically (without acceleration) and becomes in the first energy band \cite{CladePRA06}:
\begin{equation}
|\Psi_{0, q_0}\rangle=\sum_l{\phi_0(q_0+2lk)|q_0+2lk\rangle} \label{momentum space}
\end{equation}
with $l\in \mathbb{Z}$. Here $|q_0\rangle$ designs the wave function associated to a plane wave of momentum $q_0$ and the amplitudes $\phi_0$ correspond to the Wannier function  \cite{Wannier} in momentum space of the first band. When the potential depth is close to zero, the limit of the Wannier function $\phi_0$ is 1 over the first Brillouin zone and zero outside. On the contrary, if the potential depth is large, the Wannier function selects several components in the velocity space. When the optical lattice is accelerated adiabatically, the Wannier function is continuously shifted in the momentum space following the relation:
\begin{equation}
|\Psi(t)\rangle=\sum_{l}\phi_{0}(q_0 + 2lk - mv(t)/\hbar) |q_{0}+2lk\rangle\\
\label{wannier1}
\end{equation}
where $v(t)$ is the velocity of the optical lattice. The enveloping Wannier function $\phi_{0}$ is shifted by $mv(t)$ in momentum space. After the acceleration, the potential depth is decreased adiabatically and, in equation \ref{wannier1}, the Wannier function selects only one component of the velocity distribution. At the end, the wave function is $ |\Psi\rangle=|q_{0}+2Nk\rangle$. If $\Delta v$ is the velocity variation due to the acceleration, the number of Bloch oscillations $N$ is such as $|\hbar q_0+m\Delta v-2N\hbar k|<\hbar k$. Consequently, if the initial atomic velocity distribution fits the first Brillouin zone, it is exactly shifted by $2Nv_r$ without deformation. The second pair of $\pi/2$ laser pulses is used to readout the accelerated velocity distribution. By scanning the frequency of the second pair of $\pi/2$ pulses, we realize the convolution of this distribution by a second Ramsey fringe pattern. The shift of the resonance is $2Nkv_r$. The resolution of the recoil measurement is inversely proportional to the spacing between Ramsey fringes and proportional to the number $2N$ of transferred recoils.

The experimental set-up has been described previously \cite{{CladePRL},{CladePRA06}}. A cold atomic sample of $3\times 10^7$ atoms ($^{87}$Rb) is produced in a $\sigma^+ - \sigma^-$ optical molasses loaded from a magneto-optical trap (MOT). The atoms are then optically pumped to the $F=2, m_F=0$ sub-level. The optical lattice is derived from the interference of two counter-propagating vertical beams generated by a Ti-Sapphire laser. Its frequency is blue detuned by $\sim 40$~GHz from the D2 line. The lattice is adiabatically raised during 500~$\mu$s, then accelerated during 3~ms and adiabatically lowered (500~$\mu$s). The lattice depth is about $ 100~E_r$ ($E_r=\hbar^2 k^2/2m$ is the recoil energy). The $\pi/2$ laser pulses are derived from two external cavity phase-locked diodes lasers. The power of these Raman beams is $13$~mW (per beam), their radius at $1/e^2$ is 2~mm. Their frequencies are blue shifted by 310~GHz from the D2 line in order to reduce photon scattering and light shifts. Their beat frequency is precisely controlled: a frequency sweep is used during each pair of $\pi/2$ pulses in order to balance the acceleration of gravity and a frequency jump  allows to compensate the Doppler shift due to any velocity variation $\Delta v$ between the two pairs of $\pi/2$ pulses. We use a symmetric Ramsey-Bord\'e interferometer \textit{i.e.} the laser beams directions are the same for the four pulses. In this case, we are only sensitive to the velocity variation induced by the BO and the acceleration of gravity.

When we increase the number of BO up to 50, the atoms collide with the window of the vacuum chamber. To overcome this problem, we load the atom interferometer with atoms initially accelerated using also BO. In such a way, the atomic velocity  at the end of the interferometric trajectory is close to zero \cite{{CladePRL},{CladePRA06}}.

\begin{figure}
\centering
\includegraphics[width=\linewidth]{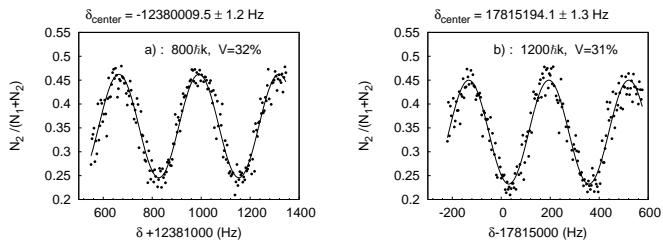}
 \caption {Typical set of two spectra used for a determination of $\alpha$ corresponding to the up (a) and down (b) trajectories. Here $N_1$ and $N_2$ are respectively the number of atoms in F=1 and F=2 after the second pair of $\pi/2$ laser pulses.}
\label{FrangeRamsey}
\end{figure}

The ratio of $h/m_{Rb}$ is deduced from four spectra obtained by using a first experimental protocol P1. It consists in : (i)
accelerating the atoms either upwards or downwards keeping the same spacing time $T$ between the two $\pi/2$ pulses pairs (the numbers of BO are respectively labeled $N^{up}$ and $N^{down}$). In both cases the velocity variation due to the gravity acceleration $g$ is $g\times T$, and can be canceled by making the difference between the two measurements. (ii) For each initial acceleration, we record two spectra by exchanging the directions of the Raman beams. In this case the contribution of some systematics arising from parasitic level shifts (ac stark or second order Zeeman effect) changes sign and is then substantially reduced by averaging these two measurements. Two typical interference fringes patterns are shown in Fig.~\ref{FrangeRamsey}. The effective number of BO is $N^{up}+N^{down}= 1000$, corresponding to 2000 recoil velocities between the upper and lower trajectories. The spacing time $T_R$ is 2.6~ms and the duration of each $\pi/2$ pulse is 400~$\mu$s.
The central fringe is determined with an uncertainty lower than $1.4$~Hz ($\sim v_r/10000$). Each spectrum is plotted with 200
points and is obtained in $6$~min. We achieve an excellent fringes visibility of about 30$\%$ for 600 BO.

\begin{figure}
\centering
\includegraphics[width=.95\linewidth]{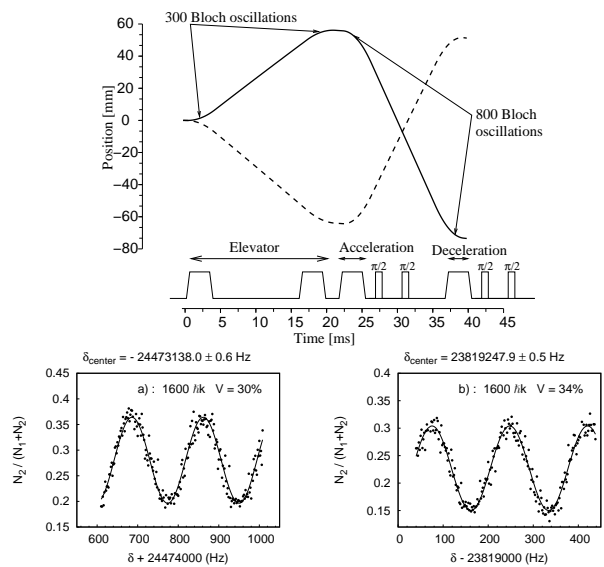}
\caption{\label{Fig:Ascenseur}Trajectories of the atom using the ``atomic elevator'' technique
and the corresponding pulses sequence. The spectra (a) and (b) correspond to an up and a down trajectory.}
\end{figure}

The number of BO and the spacing time between the two $\pi/2$ laser pulses can be potentially increased. They are still limited respectively by the displacement of the atoms in the vacuum chamber and vibrations. To overcome the first limit, we have built an ``atomic elevator''. The idea consists in displacing the atoms to the bottom or the top of the cell after the molasses phase.
For that we take advantages of the high efficiency and the precise acceleration control of BO (through the control of the frequency
 difference between the two laser beams).  We then start the measurement protocol at this position instead of the center of the cell. Fig.~\ref{Fig:Ascenseur} illustrates the new experimental sequence (protocol P2) and the corresponding atomic trajectories:
after the cooling process all the atoms are accelerated for instance in the upward direction by using  $N_{\mathrm{elv}}$ BO, thereafter when they come near the upper window they are stopped by using $-N_{\mathrm{elv}}$ BO. We then perform a measurement following the protocol P1 on a longer interaction distance. In such a way, we transfer $1600$ photon momenta and obtain a fringes pattern with a typical visibility of 30$\%$ (see Fig.~\ref{Fig:Ascenseur}). Moreover the uncertainty on the fringes center is reduced to less than 0.7 Hz by increasing the delay $T_R$ to 5.4~ms and by implementing a vibration isolation platform.

We have used both protocols P1 and P2 to obtain 221 determinations of $\alpha$. For these measurements, the effective number of BO $N^{up}+N^{down}$ varies between 200 and 1600. The dispersion of the values is $\chi^2/(n-1)$=1.85 and the resulting statistical uncertainty on $\alpha$ is $3\times 10^{-9}$.

\begin{figure}
\centering
\includegraphics[width=.95\linewidth]{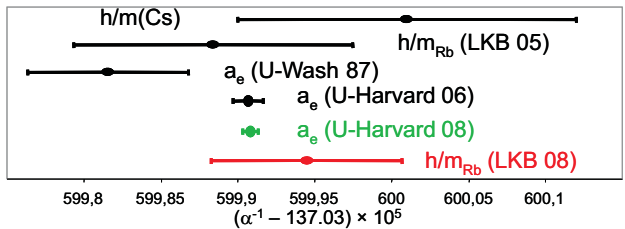}
\caption{(Color online) The most accurate determinations of the fine structure constant, in green the recent determination
deduced from the $g-2$ measurement (U-Harvard 08) and in red this work (LKB 08).}
\label{fig:Alpha}
\end{figure}

\begin{table}
\caption{\label{BudgetError} Error budget on the determination of
$1/ \alpha$ (systematic effect and relative uncertainty in ppb). }
\begin{ruledtabular}
\begin{tabular}{lcc}
\multicolumn{1}{l}{Source}
&Correction  &\parbox{2cm}{Relative uncertainty }\\
\hline Laser frequencies& &0.4\\
Beams alignment&-2& 2\\
Wavefront curvature and Gouy phase&-11.9 & 2.5\\
2nd order Zeeman effect&4.9 & 1 \\
Quadratic magnetic force&-0.59 & 0.2\\
Gravity gradient&-0.07& $0.02$ \\
light shift (one photon transition)& & 0.1\\
light shift (two photon transition)& & 0.01 \\
light shift (Bloch oscillation)&0.48 & 0.2 \\
Index of refraction atomic cloud& &0.3 \\
Index of refraction background vapor&-0.36 & 0.3 \\
Rydberg constant and mass ratio \cite{codata06}& &0.23 \\ \hline \hline
Global systematic effects&-9.54 & 3.4\\
\end{tabular}
\end{ruledtabular}
\end{table}
The systematics effects are already detailed in \cite{{CladePRL},{CladePRA06}}. For instance, in the case of the P2 protocol, they are listed in Table \ref{BudgetError}. These corrections and uncertainties are very similar to those of the 2005 measurement. We have reduced the uncertainties of the two main corrections, due to the second-order Zeemann effect and to the geometry of the laser beam, thanks to a careful mapping of the magnetic field and a precise measurement of the wavefront curvature. Finally the global correction on $\alpha^{-1}$ is $(-9.54 \pm3.4) \times10^{-9}$. Taking into account the statistical uncertainty ($3\times10^{-9}$), the total relative uncertainty on $\alpha$ is then $4.6 \times 10^{-9}$. We obtain for $h/m_{\mathrm{Rb}}$ the value $4.591\,359\,246\,(42)\times 10^{-9} ~ \rm{m}^2\cdot \rm{s}^{-1}$ and for $\alpha$:
\begin{equation}
\alpha^{-1}(\mathrm{LKB}~08)=
137.035\,999\,45\,(62)~~~~[4.6\times10^{-9}]
\end{equation}
This value is reported on the Fig. \ref{fig:Alpha} (dot labeled LKB~08) which summarizes the most relevant determinations of the fine structure constant \cite{{codata06},{VanDick},{Gabrielse2008},{Kinoshita},{Wicht},{CladePRL},{CladePRA06}}. This new determination is in agreement with our previous value (dot labeled LKB~05) and with the last result deduced from the electron anomaly \cite{Gabrielse2008}:
\begin{equation}
\alpha^{-1}(a_e)= 137.035~999~084~(51)~~~~[0.37\times10^{-9}]
\end{equation}

Although the uncertainty of this value is more than 10 times smaller than our result, the comparison of these two results provides the most stringent test of the QED. To improve it, we are investigating deeply the limits of our experiment. We are building a larger vacuum chamber to increase the number of BO. A relative uncertainty lower than $10^{-9}$ can be reasonably expected with a better control of the magnetic field and of the geometric parameters of the laser beams. This performance would provide the possibility to reach an unprecedented test of the QED theory, or, if we consider the QED theory exact, a test of a possible substructure of the electron \cite{{Odom06},{Brodsky}}.

In conclusion, we have demonstrated the possibility to combine the BO with an atom interferometer. Compared with our previous work, this combination allows to increase the resolution without decreasing the signal to noise ratio. We have obtained a new determination of the fine structure constant in agreement with our last determination based on a non interferometric method. The comparison of this result with the value of $\alpha$ deduced from the measurement of the electron anomaly is the best test of QED.
This paper also opens a new way for the design of atom interferometers based on BO. The combination of atom interferometry with BO can be used for the measurement of the gravity constant $g$ as proposed in \cite{EPL06}. It could significantly improve the sensitivity of measurements of $g$ done with a very large number of BO \cite{{Tino2006},{Naegerl}}. Thanks to a larger vacuum chamber, we plan to increase the number of BO in order to reduce the uncertainty of our measurement to 1 ppb. In this letter, we have studied a simple scheme where the BO are inserted between the two pairs of $\pi/2$ pulses. Another way to improve the precision of atom interferometers consists in using $n$-photons momentum beam splitter \cite{Chu08}. As described in \cite{Demschlag}, BO can be used to realize a large momentum beam splitter. The idea consists to initially create two coherent wave-packets by using a $\pi/2$ Raman pulse and then load those wave-packets in the first and third band of an accelerated optical lattice. By choosing suitable lattice parameters, only atoms in the first band will be accelerated, increasing the relative momentum between the two interfering paths. A preliminary calculation of the phase shifts due to the BO shows the possibility of increase the sensitivity by a factor $n$ of the order of ten.
\begin{acknowledgments}
 This experiment is supported in part by the Laboratoire National de M\'etrologie et d'Essais (Contrat 033006), by IFRAF (Institut Francilien de Recherches sur les Atomes Froids) and by the Agence Nationale pour la Recherche, FISCOM Project-(ANR-06-BLAN-0192).
\end{acknowledgments}

\end{document}